\documentclass[sigconf]{acmart}

\AtBeginDocument{%
  \providecommand\BibTeX{{%
    \normalfont B\kern-0.5em{\scshape i\kern-0.25em b}\kern-0.8em\TeX}}}

\setcopyright{acmcopyright}
\copyrightyear{2022}
\acmYear{2022}
\acmDOI{10.1145/nnnnnnn.nnnnnnn}

\acmConference[WSDM’22]{Proceedings of the ACM
International WSDM Conference}{February,
2022}{Phoenix, USA}

\acmPrice{15.00}
\acmISBN{978-1-4503-XXXX-X/18/06}

\begin{document}

\title{An Effective Way for Cross-Market Recommendation with Hybrid Pre-Ranking and Ranking Models}
\subtitle{The first-place entry for Cross-market Recommendation at WSDM Cup 2022}



\author{Qi Zhang}
\affiliation{%
  \institution{Interactive Entertainment Group of Netease Inc.}
  \city{Guangzhou}
  \country{China}
}
\email{zhangqi21@corp.netease.com}

\author{Zijian Yang}
\affiliation{%
  \institution{Interactive Entertainment Group of Netease Inc.}
  \city{Guangzhou}
  \country{China}
}
\email{yangzijian@corp.netease.com}

\author{Yilun Huang}
\affiliation{%
  \institution{Interactive Entertainment Group of Netease Inc.}
  \city{Guangzhou}
  \country{China}
}
\email{huangyilun@corp.netease.com}

\author{Jiarong He}
\authornote{Corresponding author}
\affiliation{%
  \institution{Interactive Entertainment Group of Netease Inc.}
  \city{Guangzhou}
  \country{China}
}
\email{gzhejiarong@corp.netease.com}

\author{Lixiang Wang}
\affiliation{%
  \institution{Southeast University}
  \city{Nanjing}
  \country{China}
}
\email{220191639@seu.edu.cn}

\begin{abstract}
The Cross-Market Recommendation task of WSDM CUP 2022 is about finding solutions to improve individual recommendation systems in resource-scarce target markets by leveraging data from similar high-resource source markets. Finally, our team OPDAI won the first place with NDCG@10 score of 0.6773 on the leaderboard. \footnote{The code is available at https://github.com/opdai/wsdm2022-xmrec-top1-solution.} Our solution to this task will be detailed in this paper. To better transform information from source markets to target markets, we adopt two stages of ranking. In pre-ranking stage, we adopt diverse pre-ranking methods or models to do feature generation. After elaborate feature analysis and feature selection, we train LightGBM with 10-fold bagging to do the final ranking. 
\end{abstract}



\maketitle

\section{Introduction}
E-commerce companies often operate across markets in different regions or countries around the world. How to leverage data from other markets to optimize the recommender system in a target market, namely Cross-Market Recommendation (CMR), becomes a novel and valuable topic in the industry \cite{DBLP:journals/corr/abs-2109-05929}. In this WSDM Cup challenge, we participants are provided with user purchase and rating data from various markets, with a considerable number of shared item subsets. For online validation, we need to submit a sorted list of 100 candidate items for each user in valid and test sets of 2 target markets. The evaluation metric is weighted NDCG@10 in test sets of the 2 target markets. The rest of the paper is organized as follows. We first analyze the given dataset in Section 2. Section 3 describes the details of our solution for the challenge. Experiments are illustrated in Section 4. Finally, we conclude our work and discuss the future direction in Section 5. 

\section{Exploratory Data Analysis}

\begin{figure*}
    \includegraphics[width=\textwidth]{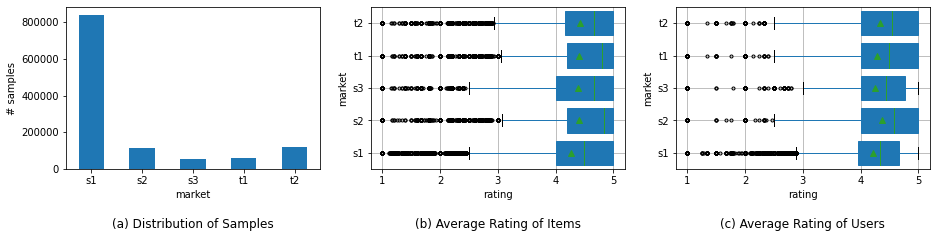}
    \caption{Distributions of samples, items' rating and users' rating per market}
    \label{fig:fig1}
\end{figure*}


\begin{table}
  \caption{General Description of Data}
  \begin{tabular}{lc}
    \toprule
    \# markets & 5 \\
    \# samples & 1147289 \\
    \# users & 196951 \\
    \# items & 116207 \\
    \# unique items & 34740 \\
    \bottomrule
  \end{tabular}
\end{table}

The competition organizer provides user-item-rating dataset from 5 markets, including 3 source markets (\textit{s1}, \textit{s2} and \textit{s3}) and 2 target markets (\textit{t1} and \textit{t2}). Additionally, the dataset in each market consists of several parts which are \textit{train\_5core}, valid and test. Basic characteristics of the dataset are shown in Table 1 and Figure 1. Source market \textit{s1} has much more samples than other markets, which may contain abundant information to boost our models. And the distributions of ratings which are mostly between 4 to 5 with average of about 4.6 are quite similar between markets, so that knowledge transfer directly based on ratings is probably reasonable. 

The set of users in each market is mutually disjoint. However, items overlapped across markets are predominate in both target markets as shown in Table 2. So building a recommender system on target markets that makes better use of items' information from other markets is quite important for this task from our perspective. \textit{s1} is a high-resource market and almost contains all items in \textit{t1} and \textit{t2}. Particularly, the number of overlapped items between \textit{t2} and \textit{s1} are the biggest among others, so it's significant to transfer knowledge from \textit{s1} to \textit{t2}. How to better utilize the data and transfer knowledge from source markets is our major focus.

\begin{table}
  \caption{Summary of Overlap Items}
  \begin{tabular}{ccccccc}
    \toprule
    Market & Total & \textit{s1} & \textit{s2} & \textit{s3} & \textit{t1} & \textit{t2}\\
    \midrule
    \textit{t1} & 3429 & 3412 & 2634 & 2007 & - & 2037 \\
    \textit{t2} & 8834 & 8782 & 2733 & 1808 & 2037 & - \\
    \bottomrule
  \end{tabular}
\end{table}

\section{Methodology}

\begin{figure}
    \includegraphics[width=\linewidth]{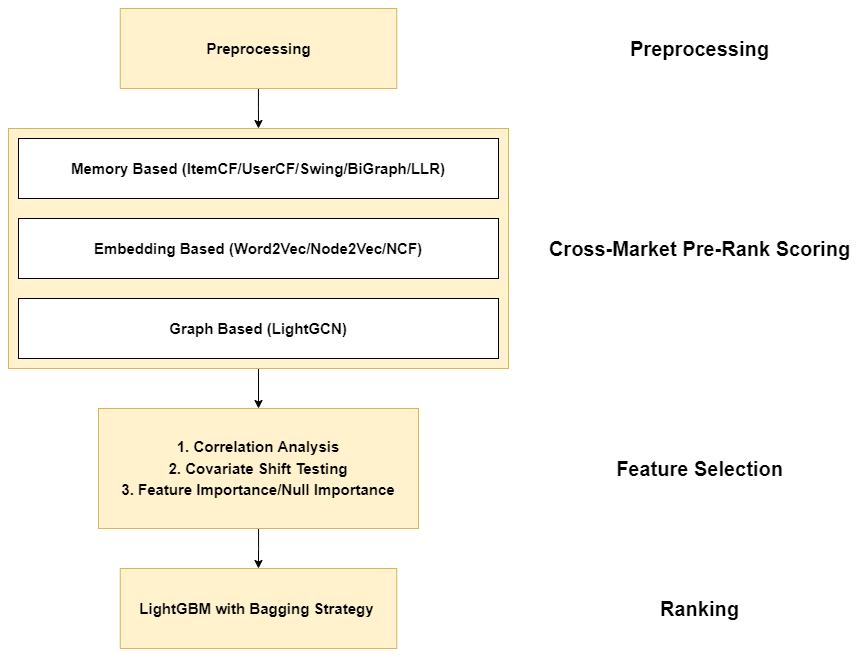}
    \caption{An overall framework and pipeline of our solution}
    \label{fig:fig2}
\end{figure}

Our solution for this task mainly consists of 4 steps, preprocessing, pre-rank scoring with cross-market data, feature selection and final ranking. This two stage training pipeline is very effective based on our experiment. The overall framework is shown in Figure 2. 

Given that the dataset provided is of high quality, we do not do much preprocessing, only label encoding, dropping duplicates and marking scores in \textit{train\_5core} set as 5 for all when models using cosine function to measure the similarities. After preprocessing, we separate model training into two stages, pre-ranking and ranking. Pre-ranking stage is not for candidates generation but for getting diverse similarity scores between users and items which are used as features in the ranking stage afterward. Therefore, pre-rank scoring can be regarded as kind of pretraining as well. However, due to the different degrees of similarities between different markets, some of the features generated through pre-ranking stage could probably be redundant or deficient in some way for one of or both of the target markets. So we do feature selection before final ranking.


\begin{table}
  \caption{Dataset using in different stages}
  \begin{tabular}{p{2.7cm}p{5cm}}
    \toprule
    Stage & Dataset \\
    \midrule
        Pre-Rank Scoring & All datasets in source markets + \textit{train} \& \textit{train\_5core} \& cross-market \textit{valid} in the target market \\
        Final Ranking Model & \textit{valid} of each target market \\
    \bottomrule
  \end{tabular}
\end{table}

As shown in Table 3 below, we use different parts of dataset in different stages. In pre-ranking stage, all source market dataset is used, besides, for target markets, we exclude the \textit{valid\_run} set individually when doing pre-rank scoring for the corresponding target market to avoid label leakage in final ranking stage. For example, when doing training for \textit{t1}, we exclude the \textit{valid\_run} dataset in \textit{t1}, using all of the \textit{train} and \textit{train\_5core} dataset and \textit{valid\_run} dataset in \textit{t2}. \textit{t2} training is following the same way.

\subsection{Pre-rank Scoring}

In pre-ranking stage, we use several different methods or models to get user/item representations and user-to-item similarity scores which represent that how much the user is interested in the given item. Based on pre-rank scoring, hundreds of diverse features are generated for the next stage of model training, which boost our final model significantly. 

To be specific, the pre-ranking models can be classified into 3 categories, which are memory-based, embedding-based and graph-based models. For memory-based models, we use some traditional collaborative filtering (CF) models like ItemCF \cite{Linden2003AmazoncomRI}, UserCF \cite{10.1145/192844.192905}, Swing \cite{DBLP:journals/corr/abs-2010-05525}, Loglikelihood Ratio (LLR) \cite{10.5555/972450.972454}, Bi-Graph \cite{10.1103/PhysRevE.76.046115} to get user-to-item similarity scores. These memory-based models above score 0.59-0.62 (NDCG@10, the same below) on the leaderboard. Especially for ItemCF, which takes the least time to train and performs the best among all of the memory-based models above. 

As for embedding-based models, Word2Vec, Node2Vec \cite{DBLP:journals/corr/GroverL16} (both in DFS and BFS ways) and NCF \cite{DBLP:journals/corr/abs-1708-05031} are used to generate user/item embeddings and user-to-item similarity scores. NCF perform the best among embedding-based models with the score of 0.61-0.62 on the leaderboard. Scores of the rest models are ranging from 0.35 to 0.46, which do not seem good enough comparing others. These models are probably not strong solo players, but also contributing to the final ranking models in some degree based on our experiment. 

Node2Vec is a simple but efficient embedding-based model. Unlike Word2Vec and DeepWalk, Node2Vec uses a biased random walk procedure to efficiently explore diverse neighborhoods in DFS or BFS ways, and thus generate richer representations. 

Graph-based models are really prevalent these years, and have become new state-of-the-art for collaborative filtering. To better represent users and items, we also adopt LightGCN \cite{DBLP:journals/corr/abs-2002-02126} to do pre-rank scoring. Specifically, LightGCN learns user and item embeddings by linearly propagating them on the user-item interaction graph. For better performance, we use 4-layer LightGCN with the embedding dimension of 2048, nodes dropout rate=0.4 and learning rate=0.001.

Moreover, considering the cross-market differentiation for user/item representations and user-to-item interest quantification, we calculate the scores based on different market combinations. For example, to generate pre-ranking scores for \textit{t1}, we can use market combinations of \textit{s1}-\textit{t1}, \textit{s1}-\textit{s2}-\textit{s3}-\textit{t1}, \textit{s1}-\textit{s2}-\textit{t1}, \textit{s1}-\textit{s3}-\textit{t1}, \textit{s1}-\textit{s2}-\textit{s3}-\textit{t1}-\textit{t2}, etc. as pretraining corpus, likewise for \textit{t2}. However, LightGCN training is relatively time-consuming, so we only conduct some of the combinations for \textit{t2} later to get the final boosting. We use ItemCF as the example to show the results of this cross-market combination strategy in Table 4. All scores are offline NDCG@10 calculated with valid set. We can see that 10 features are generated through this process for each pre-ranking model. In most cases, our models perform better with more data. However, we can find that although the data in \textit{s1} is richer than that in \textit{s3}, \textit{s3} still contributes more to \textit{t1} than \textit{s1} does with the scores 0.6805 versus 0.6781. In addition, it's obvious that different source markets have different contributions to the two target markets. \textit{s3} contributes the most to \textit{t1}, then \textit{s2} and \textit{s1} come after. As for \textit{t2}, \textit{s1} dominates the main contributions, thus \textit{s3} and \textit{s2} provide limit gains. The two target markets share different characteristics. By calculating pearson correlation coefficients between the 10 pre-rank scoring features generated through different market combinations, it's found that  the pearson correlation coefficients in \textit{t2} are much higher and of less difference between each other than that in \textit{t1}, as shown in Figure 3. Obviously, \textit{t2} market suffers serious multicollinearity problem when training the model with these features. Feature refining should be adopted to get a more precise and generalized model.

\begin{table}[!ht]
    \caption{Result of ItemCF model with cross-market combination (all scores are offline NDCG@10 in valid set)}
    \centering
    \begin{tabular}{|p{0.8cm}|p{1.5cm}|p{1.5cm}|p{0.8cm}|p{0.8cm}|p{0.8cm}|}
    \hline
        ID & \textit{t1} Market Combinations (Num. of Rows) & \textit{t2} Market Combinations (Num. of Rows) & \textit{t1} score & \textit{t2} score & \textit{t1}-\textit{t2} score \\ \hline
        \textit{s1}-\textit{s2}-\textit{s3}-\textit{t1}-\textit{t2} & \textit{s1}-\textit{s2}-\textit{s3}-\textit{t1}-\textit{t2} (1182345) & \textit{s1}-\textit{s2}-\textit{s3}-\textit{t1}-\textit{t2} (1179560) & 0.6843 & 0.5797 & 0.6142 \\ \hline
        \textit{s1}-\textit{s2}-\textit{s3} & \textit{s1}-\textit{s2}-\textit{s3}-\textit{t1} (1056685) & \textit{s1}-\textit{s2}-\textit{s3}-\textit{t2} (1116463) & 0.6850 & 0.5795 & 0.6143 \\ \hline
        s0 & \textit{t1} (60400) & \textit{t2} (120178) & 0.6776 & 0.5589 & 0.5980 \\ \hline
        \textit{t1}-\textit{t2} & \textit{t1}-\textit{t2} (186060) & \textit{t1}-\textit{t2} (183275) & 0.6789 & 0.5596 & 0.5989 \\ \hline
        \textit{s1}-\textit{s3} & \textit{s1}-\textit{s3}-\textit{t1} (926608) & \textit{s1}-\textit{s3}-\textit{t2} (986386) & 0.6839 & 0.5793 & 0.6138 \\ \hline
        \textit{s1}-\textit{s2} & \textit{s1}-\textit{s2}-\textit{t1} (999572) & \textit{s1}-\textit{s2}-\textit{t2} (1059350) & 0.6786 & 0.5793 & 0.6121 \\ \hline
        \textit{s2}-\textit{s3} & \textit{s2}-\textit{s3}-\textit{t1} (247590) & \textit{s2}-\textit{s3}-\textit{t2} (307368) & 0.6847 & 0.5604 & 0.6014 \\ \hline
        \textit{s1} & \textit{s1}-\textit{t1} (869495) & \textit{s1}-\textit{t2} (929273) & 0.6781 & 0.5783 & 0.6112 \\ \hline
        \textit{s2} & \textit{s2}-\textit{t1} (190477) & \textit{s2}-\textit{t2} (250255) & 0.6789 & 0.5601 & 0.5992 \\ \hline
        \textit{s3} & \textit{s3}-\textit{t1} (117513) & \textit{s3}-\textit{t2} (177291) & 0.6805 & 0.5606 & 0.6002 \\ \hline
    \end{tabular}
\end{table}

\begin{figure}
    \includegraphics[width=\linewidth]{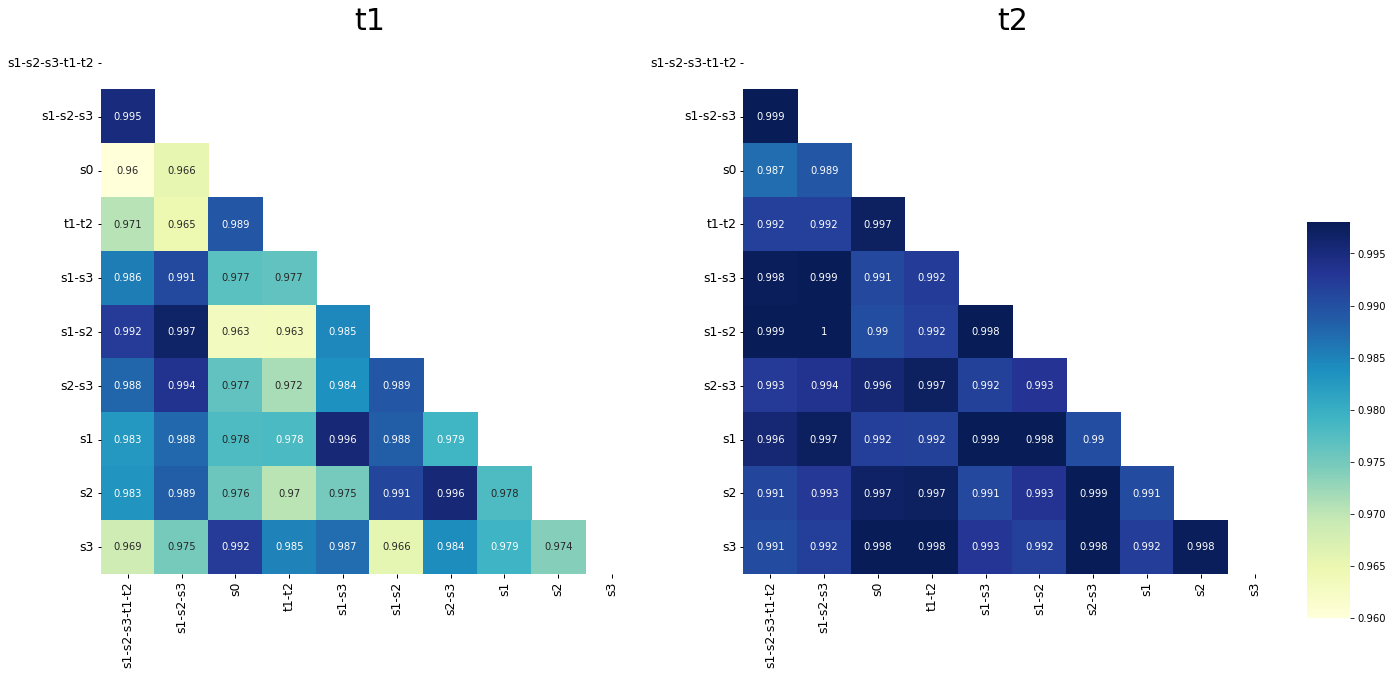}
    \caption{Pearson correlation coefficients between the 10 pre-rank scoring features generated through different cross-market combinations in \textit{t1} and \textit{t2}}
    \label{fig:fig3}
\end{figure}

\subsection{Feature Selection}

Hundreds of features are generated from pre-ranking stage, final ranking models are trained based on these features. However, many features are probably suffering distribution shifted issue between training and test set or redundant with severe multicollinearity problems as the analysis shown above. To make the final model more robust, we do further feature analysis and selection before the final model training. Our feature selection mainly based on 3 factors, which are covariate shift test \cite{U1}, offline cross-validation scores and feature importance analysis. At first, we conduct covariate shift test to exclude distribution-shift features. One of the basic and common assumptions for machine learning tasks is that  training and test dataset are independent and identically distributed. Covariate shift is a form of distribution shift between training and test set which is a significant obstacle in developing robust machine learning models. So we need to exclude features whose distributions are significantly shifted between training and test set. Specifically, we 
\balance 
refer to the method mentioned in this article \cite{U1} to do the covariate shift test. Secondly, we do heuristic feature selection based on offline k-fold cross validation scores. Under the condition of fixed seeds, folds and model hyperparameters, we eliminate a feature or a group of similar features every time to do k-fold cross validation iteratively in heuristic way. At last, we do further feature analysis based on feature importance and null importance. Null importance \cite{U2} is a very prevalent feature selection method in Kaggle. Firstly, Null importances distributions are created by fitting the models over several runs on a shuffled version of the target. And then, feature selection is adopted by comparing the null importances distributions with the actual importances gathered by fitting the models on the original target. The distance between null importances distributions and actual importances should be in a big gap and the variance of the null importances should be high if the given feature is important. After feature selection, we only keep 206 features for target market \textit{t1}, and 147 for \textit{t2}. Especially the feature selection in \textit{t2} market enables our model to get a significant boost on the leaderboard, which secure our top place on the leaderboard.

\subsection{Final Ranking}

Based on the features selected after pre-ranking stage, combining with some global statistic features, similarities calculated with pretrained Word2Vec embeddings, we build two LightGBM \cite{10.5555/3294996.3295074} classifiers to get the final ranking scores for \textit{t1} and \textit{t2} separately. 

We do not do much tuning in this stage except for some searching for model parameters \textit{num\_leaves} and \textit{learning\_rate}, which are proven to be important for the final results according to our experiments. As for model ensemble, we simply adopt bagging training with 10-fold cross validation to get a more robust model for each target market.

\section{Experiment}

With our training strategy mentioned above, we can get NDCG@10 score of 0.6737 on the leaderboard without elaborate feature selection, and we achieve our \textit{t1} sota with the score of 0.7384. 

As shown in Figure 3 above, \textit{t2} suffers from a more serious multicollinearity problem than \textit{t1} does. Meanwhile, \textit{t2} market weights more in the final score. So we did extra exploration for \textit{t2}. According to our experiment, our offline cross validation score is reliable and almost aligns with the leaderboard score, so we can use limited online submitting opportunities more efficiently by validating our ideas based on enough offline experiments. To be specific, we dropped some redundant features and optimized LightGCN for \textit{t2}  with cross-market combinations like \textit{s1}-\textit{t2}, \textit{s1}-\textit{s2}-\textit{t2}, \textit{s1}-\textit{s3}-\textit{t2}, etc., and this helps us get the final boosting from 0.6737 to 0.6773 on the leaderboard. Some related results are shown in Table 5. 

\begin{table}[!ht]
    \caption{The results of final feature optimization for \textit{t2}}
    \centering
    \begin{tabular}{|p{3.0cm}|p{1.3cm}|p{1.3cm}|p{1.3cm}|}
    \hline
        \textit{t2} features & \textit{t2} offline score & \textit{t2} online score & \textit{t1}-\textit{t2} final score \\ \hline
        Same as \textit{t1} & 0.6347 & 0.6422 & 0.6737 \\ \hline
        - Bi-Graph & 0.6348 & - & - \\ \hline
        + Node2Vec & 0.6357 & - & - \\ \hline
        + Swing & 0.6360 & - & - \\ \hline
        - User Embeddings & 0.6366 & - & - \\ \hline
        + Optimized LightGCN & 0.6378 & 0.6472 & 0.6773 \\ \hline
    \end{tabular}
\end{table}

\section{Conclusion \& Future Work}

In this paper,  we propose an effective method to boost the cross-market recommendation. To better transfer information from source markets to the target markets and avoiding biases introduced, we separate our training pipeline into two stages of ranking scoring. In pre-ranking stage, we employ diverse methods or models to do feature generation by getting pre-ranking scores. After elaborate feature analysis and feature selection, we train LightGBM with 10-fold bagging to do the final ranking individually for each target market. Finally, our team OPDAI is ranked first on the final leaderboard with the \textit{t1}-\textit{t2} NDCG@10 score of 0.6773. 

Although our approach is effective to help us get the first place in this competition, end-to-end neural networks are much more concise and flexible to solve this task in a more elegant way. We’ll leave it for future work.

\begin{acks}
XMRec organizing team paid a lot of efforts during the whole process of this competition, we really appreciate it for hosting this fantastic competition. And we would like to thank everyone associated with organizing and sponsoring the WSDM Cup 2022. 
\end{acks}


\bibliographystyle{ACM-Reference-Format}
\bibliography{sample-base}










\end{document}